\documentclass[sigconf]{acmart}

\usepackage{pifont}
\newcommand{\cmark}{\ding{51}}%
\newcommand{\xmark}{\ding{55}}%

\usepackage[nolist]{acronym}
\usepackage{scalefnt} 
\usepackage{graphicx,url}
\usepackage[english]{babel}   
\usepackage[utf8]{inputenc}
\usepackage{amsmath}
\usepackage{array}
\usepackage{algorithmic}
\usepackage{textcomp}
\usepackage{url}
\AtBeginDocument{%
  }

\acmConference[ADVANCE]{Make sure to enter the correct
  conference title from your rights confirmation email}{March 25--28,
  2026}{Brazil}
\acmBooktitle{ADVANCE: International Workshop on ADVANCEs in ICT Infrastructures and Services,
  March 25--28, 2026, Florianópolis, Brazil}

\begin{document}

\title{The Impact of Process Competition on Energy Consumption: Analysis and Modeling}

\author{Eduardo Gomes Campos}
\orcid{0009-0007-3188-7424}
\affiliation{%
  \institution{Polytechnic School of the University of São Paulo}
  \city{São Paulo}
  \state{SP}
  \country{Brazil}
}
\email{eduardogc0303@usp.br}

\author{Rafaela Sousa de Alencar Lacerda}
\orcid{0009-0001-6754-9773}
\affiliation{%
  \institution{Polytechnic School of the University of São Paulo}
  \city{São Paulo}
  \state{SP}
  \country{Brazil}
}
\email{rafaelalacerda@usp.br}

\author{Adnei Willian Donatti}
\orcid{0000-0002-4085-9640}
\affiliation{%
  \institution{Polytechnic School of the University of São Paulo}
  \city{São Paulo}
  \state{SP}
  \country{Brazil}
}
\email{adnei.donatti@usp.br}

\author{Joberto S. B. Martins}
\orcid{0000-0003-1310-9366}
\affiliation{%
 \institution{Universidade Salvador (UNIFACS)}
 \city{Salvador}
 \state{BA}
 \country{Brazil}}
 \email{joberto.martins@animaeducacao.com.br}

\author{Charles C. Miers}
\orcid{0000-0002-1976-0478}
\affiliation{%
  \institution{Santa Catarina State University}
  \city{Joinville}
  \state{SC}
  \country{Brazil}}

\author{Tereza C. M. B. Carvalho}
\orcid{0000-0002-0821-0614}
\affiliation{%
  \institution{Polytechnic School of the University of São Paulo}
  \city{São Paulo}
  \state{SP}
  \country{Brazil}
}
\email{terezacarvalho@usp.br}

\renewcommand{\shortauthors}{Tereza C. M. Carvalho et al.}

\begin{abstract}
With the development of distributed systems, the need to manage the sharing of machines among multiple simultaneous users arises. 
In the cloud computing context, the instantiation of virtual machines and containers by different users utilizing the same infrastructure leads to a dispute for physical computational resources.
In this regard, this paper analyses a process's energy consumption as a function of the competition for computational resources it encounters.
Investigating this behavior is fundamental for many applications, such as pricing in cloud computing services, and for task scheduling and load balancing, while increasing energy efficiency.
To determine this behavior, experiments were conducted and resulted in a dependency on the number of processor cores of the physical machine hosting the process.
As the number of cores increases, the process's energy consumption as a function of the competition it faces transitions from linear to a root function.
\end{abstract}




\maketitle

\section{Introduction}{\label{sec:intro}}

With the development of distributed systems, handling physical machines shared by multiple users at once is necessary. 
For instance, in the context of cloud computing, the process of instantiating virtual machines causes a scenario in which computational resources are being competed for by multiple individuals.
However, this utilization is usually measured according to the machine specifications and period of use, disregarding how each user uses those resources.
Nowadays, energy consumption and sustainability are at high stakes since the world is trying to become more aware of its environmental impact.
Since energy consumption has become a relevant cost in cloud computing data centers \cite{energyDatacenters}, there is an urge to assess the environmental impact of computer usage, and one way is to measure its energy consumption.

One topic worth mentioning is how the competition for \acp{CPU} resources affects the power consumption of a process.
In this context, this article proposes an approach to analyze the energy consumption at the process level in these scenarios, which unveils the possibility of modeling process power.
%
%
%
In this context, we perform experiments to measure how much energy a process would consume as it faces competition for CPU resources.
With that data available, we correlate how high the competition is and how much power the main process consumes.
The final objective is to obtain a function $\mathcal{W}(p)$ to represent the power consumed by the process. 
In this case, $p$ represents the percentage of the CPU used by the competition and, therefore, $p \in [0,100 - q]$, given that $q$ represents the percentage of the CPU used by the analyzed process.

To establish a reliable $\mathcal{W}(p)$ function, we comprehend diversified variables that might cause an impact on the obtained metrics.
Per \cite{experimentsmodels1}, several factors might impact the power consumption on a computer besides CPU usage. 
Despite the processor being highlighted as the main energy consumer according to \cite{CPUpower}, elements such as fans, memory, and disk usage can also have their share of effects on this topic. 
Moreover, there have been studies that modeled those variables as energy consumption \cite{compreheensiveM}. 
Thus, we replicate those tests in machines with different hardware configurations to get feasible results. 

When it comes to real-world applications, this paper suggests a model that could be used in many scenarios. 
As it tracks energy consumption at the process level, the function can estimate how much energy will be consumed by a single application running on the computer.    
For instance, in the context of cloud pricing, it is common to see the hardware selected by the user as the main definer of how much will be charged \cite{AmazonPricing}.
However, with our model, the cloud provider could differentiate how each user requires resources from the machines (by tracking its processes). 
This could lead to different and more accurate prices for each individual. 
Besides, other interesting applications might be related to large energy forecast models.
As an example, during the instantiation of \acp{VNF}, a theme of interest has been finding ways to allocate them with energy efficiency criteria \cite{vnfplace}.
With our model, it is possible to estimate how much energy a \ac{VNF} will consume given the competition for CPU resources on a certain machine.
Hence, this could be used to optimize the placement of these functions to reduce total energy consumption.

The rest of the paper is organized as follows: 
In Section~\ref{sec:problem}, we explain the problem to which our solution will be applied, clarifying on which domains our work collaborates with the state-of-the-art.
Section~\ref{sec:back} discusses the state-of-the-art models for energy consumption regarding CPU usage and delves into theoretical definitions for our applications.
Section~\ref{sec:firstpage} we bring up papers that are related to our work on topics such as resource utilization modeling in virtualized contexts, the influence of hardware on energy consumption, etc. 
These will be presented as theory background to give the necessary presumptions to set up the experiments and to justify the obtained conclusions.
Additionally, Section~\ref{sec:prop}, describes the methodology utilized to develop the experiments utilized for the model construction.
Section~\ref{sec:env} describes both the experiment environment and its main tools. 
Moreover, Section~\ref{sec:figs} showcases the results achieved with those tests graphically and discusses the thought process with each experiment's results. 
Section~\ref{sec:cons} concludes the article by summarizing the conclusions obtained from the experiments' results.


\section{Problem Definition and Motivation}
\label{sec:problem}


In this context, describing problems associated with process resource sharing is imperative. 
Among those, we highlight the instantiation of \acp{VNF} in a network-slicing context \cite{martins_enhancing_2023} \cite{donatti_energy_2025}. 
With the development of 5G mobile networks, one of its core components is \ac{NS}, which proposes a dynamic provisioning system of networking resources (and functions) to achieve compliance with user needs \cite{NS5g} \cite{NS5g2}. 
In this situation, one relevant theme of research has been the placement of \acp{VNF} to achieve high energy efficiency.
In this matter, previous work \cite{vnfplace} \cite{vnfplace2} \cite{vnfplace3} has used models to estimate energy consumption and use this information to apply statistical and algorithmic approaches to provide techniques to optimize the placement of such functions.

Nevertheless, most models do not specify the power usage of a single function but rather analyze the power dissipation of the system as a whole.
That is, we cannot analyze the consumption of each VNF and, thus, it becomes harder to estimate the energy consumption of a single slice.
Therefore, describing a process's energy consumption within the context of the entire system is essential since it would allow measuring the power dissipation of each of its components.

We focus on assessing how a process's power dissipation behaves when it faces competition for CPU resources. 
Thus, we describe it as a function $\mathcal{W}(p)$, given that $p$ stands for the competition the process is facing at a certain instant.
%
%
%
The behavior of a process's power dissipation is a key factor for VNF placement heuristics and algorithms generating a network topology that guarantees energy efficiency and balance among slices, for instance.
Therefore, this showcases the importance of such a description.

In this sense, we proceed by formulating a sample VNF placement problem described as: let $\mathcal{M}$ be a set composed of different machines that would be used to create Network Slices.
Then, $\mathcal{S}$ is a set of the created slices and $\mathcal{F}$ is a set of VNFs to be placed on these machines.
Thus, each element from $\mathcal{F}$ belongs to a certain element from $\mathcal{S}$, since, in this problem, a VNF composes a fraction of a slice running in a certain machine.

\begin{figure}[ht]
\centering
\includegraphics[width=.47\textwidth]{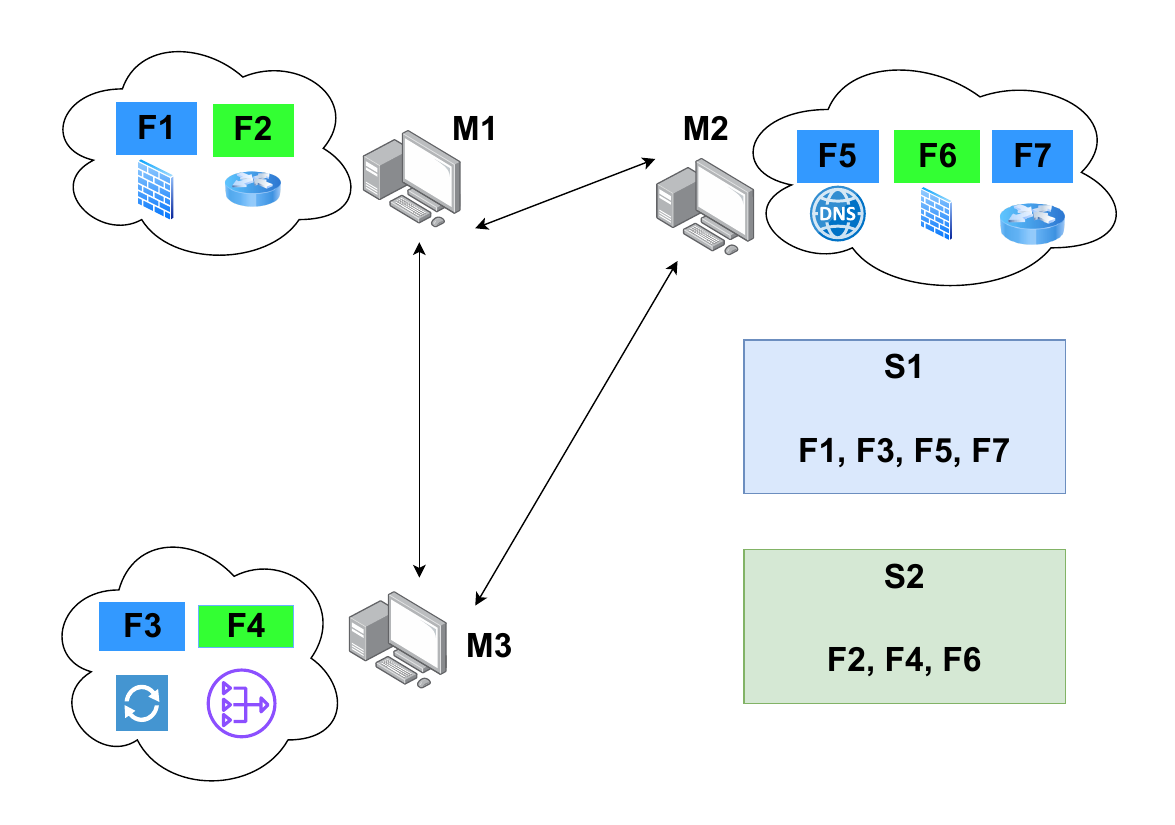}
\caption{Diagram representing a topology for the \acp{NS}}
\Description{A diagram representing a topology for the Network Slices. The figure shows two different slices composed by shared physical resources.}
\label{fig:fignet}
\end{figure}

To exemplify such a problem, we describe it with Figure ~\ref{fig:fignet}.
In this situation, we have 2 network slices (S1 and S2) built with 3 different machines (M1, M2, M3). 
Each machine hosts a few network functions (F1, F2, ..., F6, F7), and each of them either belongs to S1 or S2, as previously suggested with the problem definition.


\begin{equation} \label{eq:set}
	\mathcal{M} = [m_1, m_2, ..., m_a]
 \end{equation}
 \begin{equation}
     \mathcal{S} = [s_1, s_2, ..., s_b]
 \end{equation}
  \begin{equation}
    \mathcal{F} = [f_1, f_2, ..., f_c]
\end{equation}
Then, we define the power dissipated by a single VNF as $W(f_j)$, $j \in [1,c]$, the power dissipated by each slice as $P(s_i)$, $i \in [1,b]$, and the power dissipated by whole slice group as $P'(\mathcal{S})$:  
\begin{equation} \label{eq:powslice}
	P(s_i) = \sum_{j=1}^c q_{i,j} . W(f_j)
 \end{equation}
 \begin{equation} \label{eq:powtotal}
	P'(\mathcal{S}) = \sum_{i=1}^b P(s_i) = \sum_{i=1}^b \sum_{j=1}^c q_{i,j} . W(f_j)
 \end{equation}
 \begin{equation}
     q_{i,j} =
\begin{cases}
1 &\text{if $ f_j \in s_i$} \\
0 &\text{else}
\end{cases}
\end{equation}

As we wish to minimize the power consumption (\ref{eq:powslice}) of all the slices in $\mathcal{S}$, there is an urge to develop a description for $W(f_j)$.
Since VNFs could be initially placed in any machine, we suggest the following structure for its power consumption $W(f_j)$, given that $p_{mk}$ represents the competition for resources on a certain machine $m_k$:
\begin{equation} \label{eq:defW}
	W(f_j) = minimum[\mathcal{W}(p_{m1}),\mathcal{W}(p_{m2}),..., \mathcal{W}(p_{ma})]
\end{equation}

This means that one can choose which machine $m_k$ to allocate a VNF based on how the competition for resources in each $m_k$ affects the VNF's power consumption.
With this construction, the description we found for $\mathcal{W}(p)$ is useful in optimization problems for the placement of VNFs, considering the competition for processing resources. 
Moreover, it would allow a per-slice energy analysis even if $s_i$ shares the same physical machine with other NS. 
Thus, displaying the potential of our research. 
With this motivation, this paper aims to answer the following research questions:
\begin{itemize}
    \item Q1: How to describe Energy Consumption mathematically?
    \item Q2: How do CPU usage and energy consumption correlate?   
    \item Q3: How to estimate the energy consumption of a process?
    \item Q4: How does competition for resources affect the energy consumption of a process?
\end{itemize}
%
By the end of our research, we expect to prove or refute the following hypothesis: the energy consumption of a constant process is constant, regardless of the competition for resources it faces.


\section{Background}{\label{sec:back}} 

\textit{Energy} is the basic resource needed for performing human activities in the current century, especially in the field of computing.
However, its definition is easily misunderstood for that of \textit{Power}, and it is important to differentiate those terms.
Energy (E) represents the total work done by a system over a specified period of time (T), as shown in (\ref{eq:energy}), whereas power (P) refers to the rate at which the system performs that work \cite{energySurvey}.
Therefore, power can be defined as the infinitesimal amount of energy consumed or produced in a very small time interval, as shown in (\ref{eq:power}).
In computer systems, it is common to use power instead of energy when analysing the consumption profile.
Since power is an instantaneous quantity, it offers more precise insights into how energy consumption varies dynamically over time.

\begin{equation} \label{eq:energy}
	E = \int_{0}^{T} P(t) \, dt
\end{equation}

\begin{equation} \label{eq:power}
	P(t) = \frac{dE(t)}{dt}
\end{equation}

The energy consumption of a computer derives from several of its components, such as CPU, memory, storage, fans, and the network interface card.
However, as the CPU consumes the most energy compared to other components, it is common in power modeling to take the energy consumed by the processor as representative of the entire machine's consumption \cite{energySurvey} \cite{10.1145/1250662.1250665} \cite{studypcpower}.
Work \cite{CPUpower} also highlights that the energy consumed by memory and network devices is insignificant compared to that of the CPU.

The relationship between power consumption and CPU usage is a common subject of study in the computing field \cite{energySurvey} \cite{serverCPUen}.
In this context, several studies have been conducted over the years to determine what this relationship is.
Even though some studies propose that different functions (such as polynomial and non-linear \cite{cpupolynomial} \cite{cpuweirdnonlinear}) can characterize the relationship between energy consumption and CPU usage, it is generally accepted that this behavior could be linearly represented.
The research in \cite{CPUpower} has significantly influenced power modeling for data centers and proposes this linear correlation and also the model (\ref{eq:powerModel}) to represent it mathematically. 
In (\ref{eq:powerModel}), $P_u$ is the power consumption of the server as a function of the CPU usage, $u$.
$P_{idle}$ and $P_{max}$ are constants, the average power consumption when the server is idle and at its maximum capacity, respectively.
\begin{equation} \label{eq:powerModel}
	P_u = (P_{max} - P_{idle})u + P_{idle}
\end{equation}

Studies such as \cite{studypcpower} have shown that this linear model can accurately describe a server's power consumption. 
Therefore, as a way of simplifying hands-on measurement, this article will consider the CPU as the only component responsible for a machine's energy consumption.
We will also consider the relationship between energy consumption and CPU usage as linear, as it is classically considered by literature \cite{energySurvey} \cite{serverCPUen}.

However, this state-of-the-art relationship is only representative of the energy consumption of a computer, and cannot be used at the process level.
The survey \cite{energySurvey} extensively reviews power models and finds that only \cite{processModel} proposes a model for the energy consumption of a process.
Moreover, \cite{processModel}'s model doesn't consider the CPU usage, and does not provide a relationship between that and the energy consumption of a process.
Therefore, literature does not provide a model for the relationship between a process's energy consumption and the machine's CPU usage.
In this context, this paper aims to empirically find such a model that works at the process level.


\begin{table*}[!t]
\caption{Summary of comparison between papers}\label{tab:related_work}
\begin{tabular}{ | >{\centering\arraybackslash}m{3.5cm} | >{\centering\arraybackslash}b{2.4cm}| >{\centering\arraybackslash}b{2.4cm} | >{\centering\arraybackslash}b{2.4cm} | >{\centering\arraybackslash}b{2.4cm} | >{\centering\arraybackslash}b{2.4cm} |}
  \hline
  Research questions & \cite{docker} & \cite{experimentsmodels1} & \cite{CPUEnVms} & \cite{experimentsmodels2} & This work\\
  \hline
  Q1 & \xmark & \cmark & \xmark  & \cmark & \cmark \\
  \hline
  Q2 & \cmark & \cmark & \cmark & \cmark & \cmark\\ 
  \hline
  Q3 & \xmark & \xmark & \xmark & \xmark & \cmark\\
  \hline
  Q4 & \xmark & \xmark & \xmark & \xmark & \cmark \\ 
  \hline
\end{tabular}
\end{table*}

Virtualization technologies allow multiple activities to be run independently on the same physical machine, facilitating isolated environments that are essential for accurate performance and energy consumption analysis \cite{virtualOVRVW}.
VMs, a hypervisor-based type of virtualization, provide robust security and isolation at the hardware level \cite{containerRVW}.
However, the overhead associated with VMs can significantly impact the energy consumption profile of individual processes, thereby complicating precise measurement and analysis \cite{virtualTech}.

Containers, though not as secure, are lightweight and can be rapidly instantiated \cite{virtualTech} \cite{containerRVW}.
They also provide sufficient isolation for most application scenarios and have a negligible impact on the host machine’s energy consumption \cite{PINE}.
This makes containers particularly suitable for performance isolation and energy consumption research.
Furthermore, Kubernetes is an open source system for automating deployment, scaling, and management of containerized applications \cite{Kubernetes}.
The dynamic resource allocation it implements in containerized environments enhances the efficiency and flexibility of resource usage \cite{kubeDynamic}.
This further mitigates the overhead concerns typically associated with virtualization. 

By using containerization, we can more accurately assess the energy consumption of individual processes, ensuring that the overhead introduced by the virtualization layer remains minimal and does not distort the measurements.
Thus, we chose containerization to investigate the energy consumption of a process, benefiting from the low overhead, rapid deployment, and effective resource management capabilities inherent to this technology.

The phenomenon of two or more simultaneous processes competing for computational resources is called ``resource competition'' \cite{PINE}.
When multiple containers operate on the same physical machine, they compete for processing resources.
Thus, there is resource competition associated with processing resource sharing, and it is imperative to learn how that affects the energy consumption of each container.
However, the container's energy consumption can be broken down to the process level.
Therefore, learning the relationship between a process's energy consumption and the resource competition it encounters enables the development of generalized solutions.
These solutions would not only be applicable to containerized environments but could also extend to other process-based applications.


\section{Related Work} \label{sec:firstpage}


When looking for work related to ours, our search method consisted of searching the IEEEXplore platform with the following advanced search command, with keywords related to our research:
\textit{("All Metadata":energy consumption) AND ("All Metadata":monitoring) AND ("All Metadata":virtualization) AND (("All Metadata":cpu usage) OR ("All Metadata":energy efficiency))}.
Filtering the articles from the last 9 years, from 2015 to 2024, we found that \cite{docker} and \cite{CPUEnVms} were the resulting papers that related best to our work.
Besides, we found two other articles that relate to our work, \cite{experimentsmodels1} and \cite{experimentsmodels2}, when reviewing literature and reading through the survey \cite{energySurvey}.
This survey is relevant in the Energy Consumption Modeling field, since it thoroughly reviews models for energy consumption in data centers.
\cite{energySurvey} has 410 citations and is cited by 733 papers.

Work \cite{CPUEnVms} uses test scenarios to assess how the energy consumption of \acp{VM} behaves when sharing virtual network infrastructure and CPU cores.
Throughout the article, it evaluates how the distribution of network traffic between \acp{VM} results in different values of power being dissipated by the machines altogether.
Moreover, it also delves into how the competition for processing resources at CPU cores generates different energy consumption for the VM cluster.
Nevertheless, unlike our contribution, this paper does not analyze the power consumption at the process level, only from the point of view of the VM.

Besides, \cite{experimentsmodels1} and \cite{experimentsmodels2} provide a similar analysis of how power consumption correlates to CPU utilization.
This correlation is assessed by executing experiments in which they measure the total power consumed by the PC as it faces a gradual increase in CPU utilization over time.
Afterward, \cite{experimentsmodels1} shows how linear and non-linear empirical functions can describe the behavior of the power dissipated on similar machines.
On the other hand, \cite{experimentsmodels2} uses a polynomial function to approximate $\mathcal{W}(p)$. 
Still, differentiating from our work, the first one does not consider different hardware configurations, and neither of them assesses process-level energy consumption.

Furthermore, \cite{docker} introduces a study into the power consumption behavior of Docker containers. 
This article investigates how the containers' instantiations create energy overhead and how their power dissipation behaves when facing gradually increasing loads over time in different applications.
For example, they run containerized versions of Nginx servers and submit them for a test that increases the number of requests throughout the experiment, establishing a scenario of competition for processing resources related to our work.
Even though this paper assesses energy consumption in containers and includes CPU-sharing elements, it does not propose any process-level power representation or model.

Finally, Table~\ref{tab:related_work} presents a summary of the contributions of each paper mentioned alongside our production. 
Although the CPU use and energy consumption relation is a well-discussed topic in the literature, it remains to be introduced an analysis of process-level energy consumption regarding the total use of the CPU.
Hence, this paper proposes a study that complements previous work done on this subject.


\section{Proposal}\label{sec:prop}


To solve the problem of describing the impact of competition on the energy consumption of a reference process (a \ac{VNF}, for instance), we propose an empirical method. 
In this sense, our method consists of running various experiments and observing how the power dissipation of a specific process behaves by collecting data on certain metrics.
In the end, with extensive experimentation and data collection, we can describe the behavior of power dissipation.
Accordingly, our method relies on three pillars:

\begin{itemize}
    \item \textbf{Baseline Process}: We establish a baseline process - the baseline process has a constant workload and serves as a reference for the competition.
    \item \textbf{Observed Metrics}: 
    \begin{itemize}
        \item resource usage - we observe the resource usage (more specifically, CPU) of the machine and of the baseline process. This allows us to measure the intensity of competition through its resource usage; and
        \item power consumption per process - we gather data on the power consumption of the baseline process.
    \end{itemize}
    \item \textbf{Experimentation Scenarios}: we define scenarios in which we change variables, such as hardware (experiments are reproduced in different machines with different hardware); active cores (within the same machine); and competition growth (e.g., none, gradual increase). This allows us to investigate how each of these variables affects the observed metrics.
    
\end{itemize}

\section{Experiments environment}{\label{sec:env}}

\begin{table}[H]
\begin{center}
\resizebox{8cm}{!}{
\begin{tabular}{ | >{\centering\arraybackslash}m{1.5cm} | >{\centering\arraybackslash}b{1.5cm}| >{\centering\arraybackslash}b{1.5cm} | >{\centering\arraybackslash}b{1.5cm} | >{\centering\arraybackslash}b{1.5cm} |} 
  \hline
  Machine name & CPU & Total RAM available & Disk Space available & Number of Threads\\
  \hline
  Controller & Intel(R) Core(TM) i7-4770 & 7.67 GB & 1 TB & 8 \\ 
  \hline
  Worker 1 & Intel(R) Core(TM) i5-3330 & 15.5 GB & 1 TB & 4 \\ 
  \hline
  Worker 2 & Intel(R) Core(TM) i7-2600 & 15.5 GB & 500 GB & 8 \\
  \hline
  Worker 3 & Intel(R) Core(TM) i7-2600 & 7.63 GB & 500 GB & 8 \\
  \hline
  Worker 4 & Intel(R) Core(TM) i5-8500 & 3.65 GB & 1 TB & 6 \\
  \hline
  Worker 5 & Intel(R) Core(TM) i7-2600 & 7.63 GB & 500 GB & 8 \\
  \hline
\end{tabular}}
\end{center}
\caption{Machines configurations}\label{tab:hardware}
\end{table}

To correlate CPU competition usage and process energy consumption, it is necessary to execute tests to attempt to fit the obtained data into a mathematical model that represents this correlation. 
To achieve this, we created an environment in which we could carry out these experiments.
In this context, we assembled 6 machines (hardware configurations in the table~\ref{tab:hardware}) and used them to create a Kubernetes Cluster (Client version 1.28.1 and Server version 1.28.6). 

Creating a cluster allows automating tests on multiple machines with centralized telemetry capabilities. 
In this sense, we deploy pods running Docker images with specific test routines so that they can be run as pods on each computer.
When it comes to this cluster architecture, the machines are labeled as Controller or Worker.
The Controller is responsible for managing the cluster.
That is, it makes sure the requirements defined by the user (such as scheduling pods and guaranteeing connection between nodes) are met.
On the other hand, the Worker nodes are responsible for hosting the pods running the test routines.
To execute the tests and fetch the data obtained from them, we created a four-agent scheme: 
Scaphandre pods to extract energy information from the Worker nodes, a Prometheus agent so that this data could be exported, a Grafana agent to plot graphs with the exported data, and a Python client to interact with the Controller API.

\begin{figure}[ht]
\centering
\includegraphics[width=.47\textwidth]{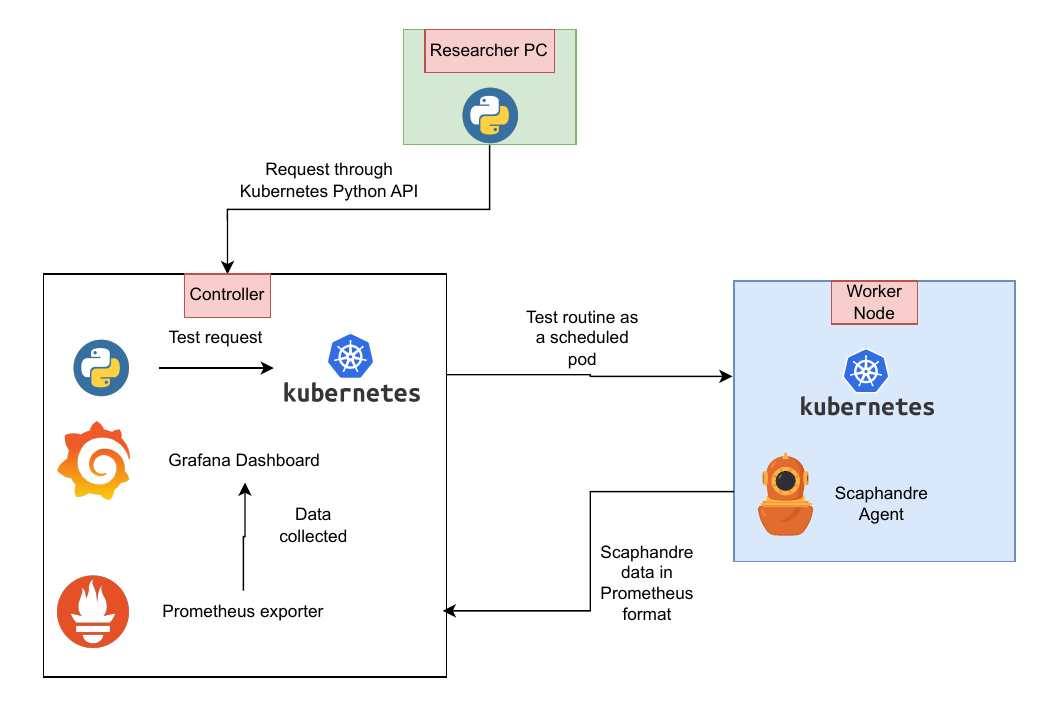}
\caption{Diagram representing the test environment and its main agents}
\Description{A diagram representing the test environment. The diagram shows elements such as the researcher station, worker node, and controller node. Each element is associated with its running technologies/tools (e.g., Grafana, Python Scripts, Kubernetes, Prometheus, Scaphandre).}
\label{fig:figdiag}
\end{figure}


Then, a tool with process-level power telemetry is needed to gather the goal data. 
Even though there have been previous software that provided these capabilities \cite{proclvl1} \cite{proclvl2}, Scaphandre \cite{scaph} stands out due to its natural compatibility with Kubernetes and Prometheus, making it more suitable for our distributed measurements. 
By using the Powercap RAPL sensors integrated into Intel CPUs (which have been previously validated and used to assess their power consumption \cite{rapl}), it can track the power dissipated by the host as a whole and estimate energy consumption by processes. 
Thus, by selecting a process on the machine (by tracking its PID, for instance) the agent can estimate how much energy the chosen element consumes, enabling the aforementioned process-level analysis.

Moreover, the Prometheus and Grafana agents act together to obtain this power information and expose it graphically on time-series dashboards (over the test duration). 
Finally, the Python client consists of a tool to interact with the controller through the Kubernetes API. Through this interaction, we could run tests remotely by authenticating and deploying the test pods.


\section{Baseline}\label{sec:baseline}
In order to validate our method, we ran two initial experiments.
In both, we analyzed the power consumption of a baseline process, but in different scenarios.
Through the command-line tool for Kubernetes, kubectl \cite{kubectl}, we were able to use the Stress tool \cite{stress} to generate 2 constant loads, that is, the baseline process and its competition.
These two tests were run on Worker 1.
Scaphandre scraped power consumption data, which was stored in Prometheus, and Grafana was used to export it to a .csv file.

\begin{figure}[!htbp]
  \centering
  \begin{minipage}[b]{0.45\textwidth}
    \includegraphics[width=\textwidth]{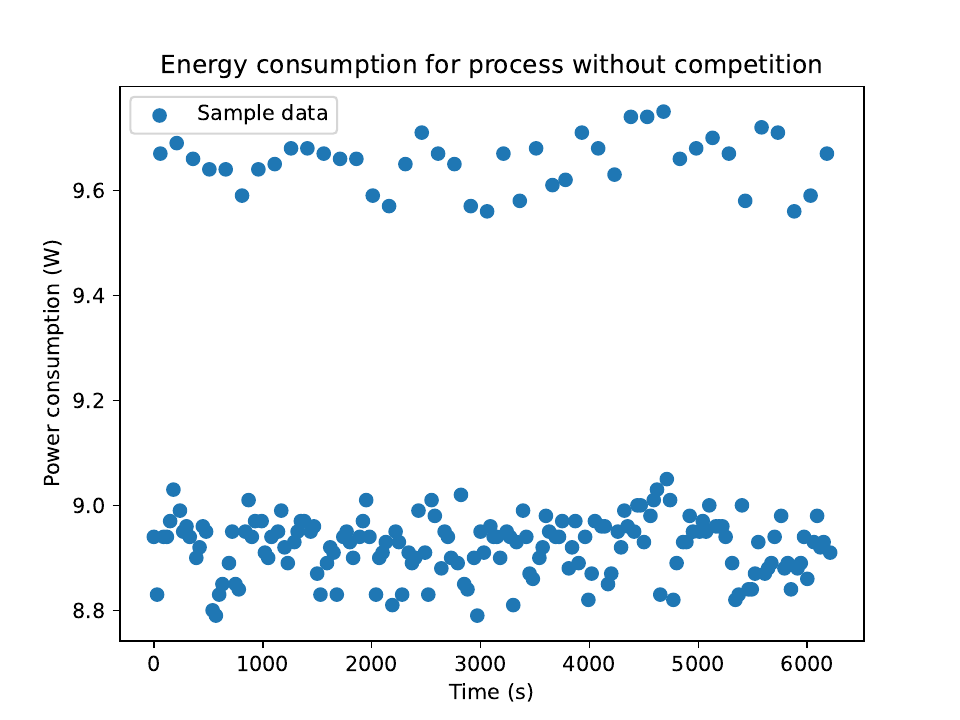}
    \caption{Power consumption of a constant process}
    \Description{Graph showing the power consumption behavior of a constant process}
    \label{fig:alone}
  \end{minipage}
  \hfill
  \begin{minipage}[b]{0.45\textwidth}
    \includegraphics[width=\textwidth]{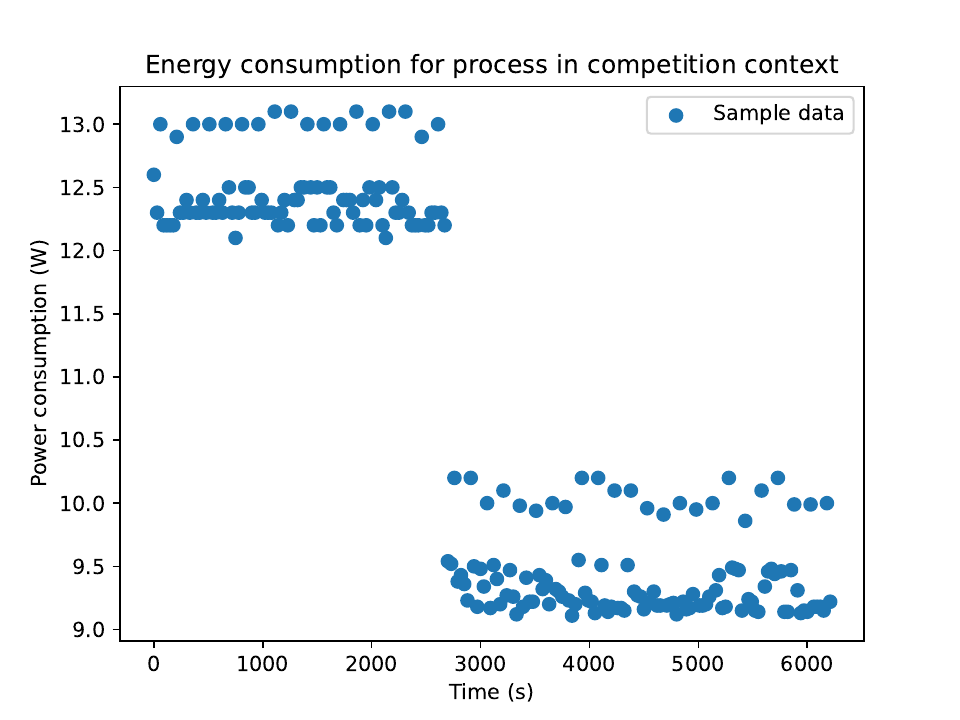}
    \caption{Power consumption with and without competition}
    \Description{Graph manifesting the differences on power consumption with and without competition}
     \label{fig:comp}
  \end{minipage}
\end{figure}

The first scenario was of no competition for resources, that is, with only the baseline process consuming resources (in this case, CPU).
We found that its power consumption (shown in Figure \ref{fig:alone}) varies from 8.7 to 9.8 Watts, which is a 11.9\% variation of the average value, proving that the consumption of the baseline process is in fact constant.
Then, the second scenario consisted of running another process alongside the baseline, and identical to it.
After some time, the second process was deactivated, enabling comparison between the first scenario (no competition) and the second (with competition).
In Figure \ref{fig:comp}, the sample on the left side, with an average value of 12.5 Watts, corresponds to Scenario 2, and the sample on the right, with an average of 9.75 Watts, corresponds to Scenario 1.

This proves that our method enables us to describe the impact competition has on the energy consumption of a reference process, through empirical means.
Furthermore, this raises the possibility that competition increases the energy consumption of a baseline process, contradicting our initial hypothesis that the energy consumption of a constant process is constant.
With this motivation, we ran more experiments in order to study the behavior of a process's energy consumption as it faces resource competition.


\section{Experiments and Results}\label{sec:figs}
\begin{figure}[ht]
\centering
\includegraphics[width=.49\textwidth]{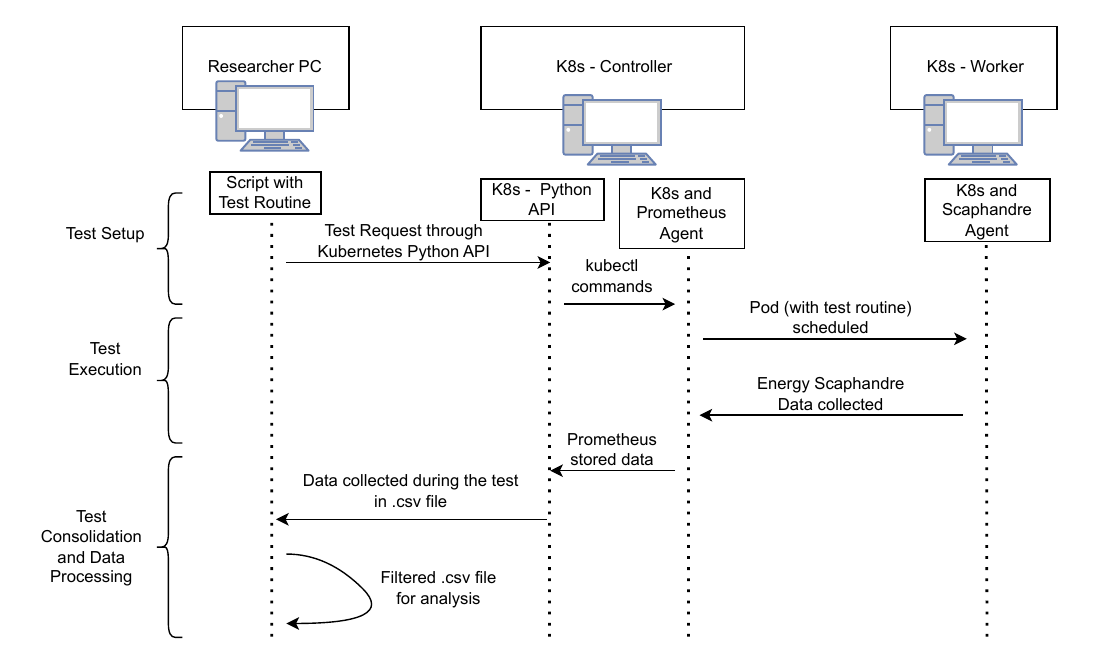}
\caption{Sequential Diagram describing the test routine}
\Description{Figure with a sequential diagram showing the test routine}
\label{fig:seqdiag}
\end{figure}

Figure \ref{fig:seqdiag} displays sequentially the tests routine. 
As Section \ref{sec:env} described, we initially code the test routine and, with the communication with the Kubernetes Python API, the Controller is able to fetch the necessary information to set up the test.
In this situation, the API, through kubectl commands, is able to create the necessary jobs and services to run the test.
Afterwards, the Kubernetes agents on the Controller and Worker nodes are able to set up the pods with the test scripts. 
Besides, they also communicate in order to get the energy data collected by Scaphandre in the Worker running the routine. 
These sets are sent to the Controller, which stores them with the Prometheus volume.
After the test is finished, the data stored in Prometheus is then compiled in a .csv file by the API, which is sent to the Researcher PC.
The information is filtered and processed with statistical analysis, giving the final results, which are discussed in this section.

To assess the competition effects on CPU energy consumption, we used both the CpuLimit \cite{cpulim} and Stress \cite{stress} tools to generate processing loads and test routines. 
While the Stress tool generated the load itself (such as running a C script in a loop), CpuLimit was responsible for controlling the load to achieve the expected percentage of usage required for the investigation.

Regarding the test routine, their main structure consisted of running a main process with a constant CPU load and gradually increasing the competition by instantiating smaller processes over time.
During the experiment, the main process is monitored, and its power consumption is exposed in Grafana dashboards.
Then, the plotted data was analyzed statistically to evaluate whether the data would fit in a specific model for $\mathcal{W}(p)$.

\subsection{Resource Competition Experiment - Gradual Increase}{\label{subsec:grad}}

Initially, we proposed the previously detailed routine: the main constant process facing escalating competition for CPU resources while getting its energy measured. 
In our investigation, we propose that the competition starts at 0\% and increases 5\% at a time every 6 minutes (to gather enough data targeting reducing the effects of outliers), until the total processor usage reaches approximately 100\%.
These cycles were repeated 8 times to reduce the effect of possible outliers and the obtained data was saved in a CSV file.
The obtained data was saved and published at Zenodo for public access\cite{campos}.
Then, by using a Python program that receives a math model from the user (such as linear, quadratic, cubic, etc.), we fit the data into the model and evaluated its correlation by using a t-test.
To get started, we executed this experiment on all Worker machines.

\begin{figure}[!htbp]
  \centering
  \begin{minipage}[b]{0.4\textwidth}
    \includegraphics[width=\textwidth]{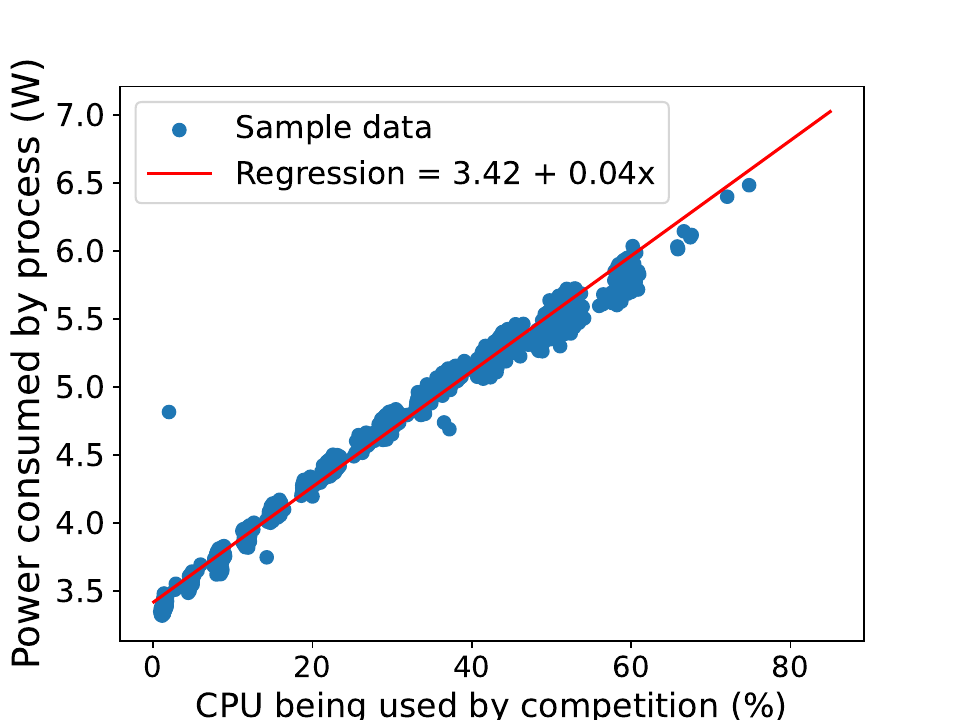}
    \caption{Worker 1 results}
    \Description{Graph showing the results for worker 1}
    \label{fig:w1}
  \end{minipage}
  \hfill
  \begin{minipage}[b]{0.4\textwidth}
    \includegraphics[width=\textwidth]{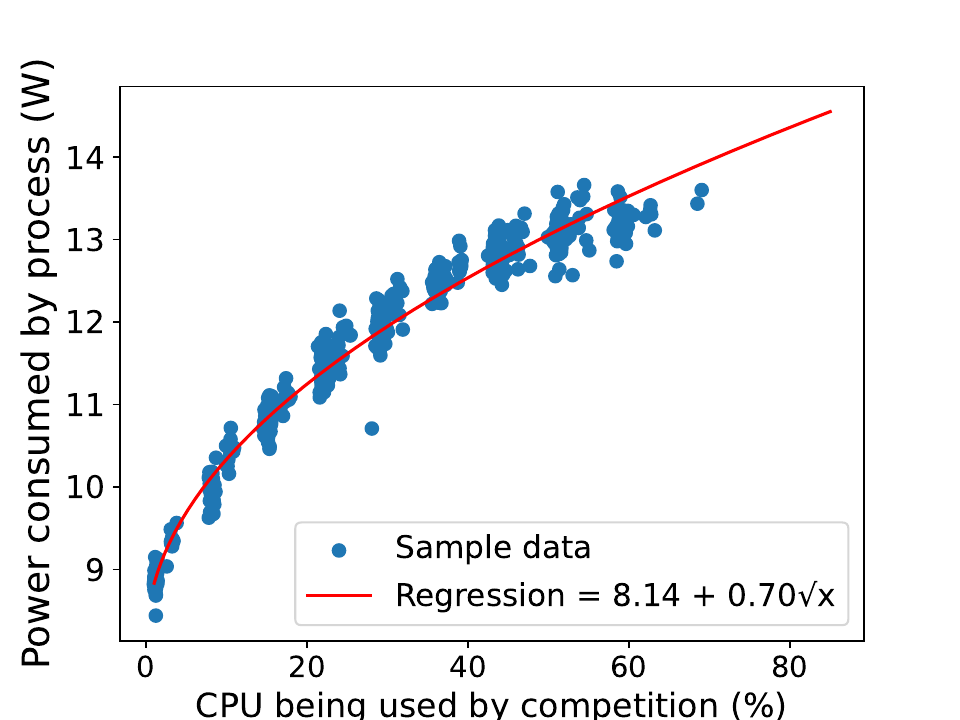}
    \caption{Worker 2 results}
     \label{fig:w2}
  \end{minipage}
\centering
  \begin{minipage}[b]{0.4\textwidth}
    \includegraphics[width=\textwidth]{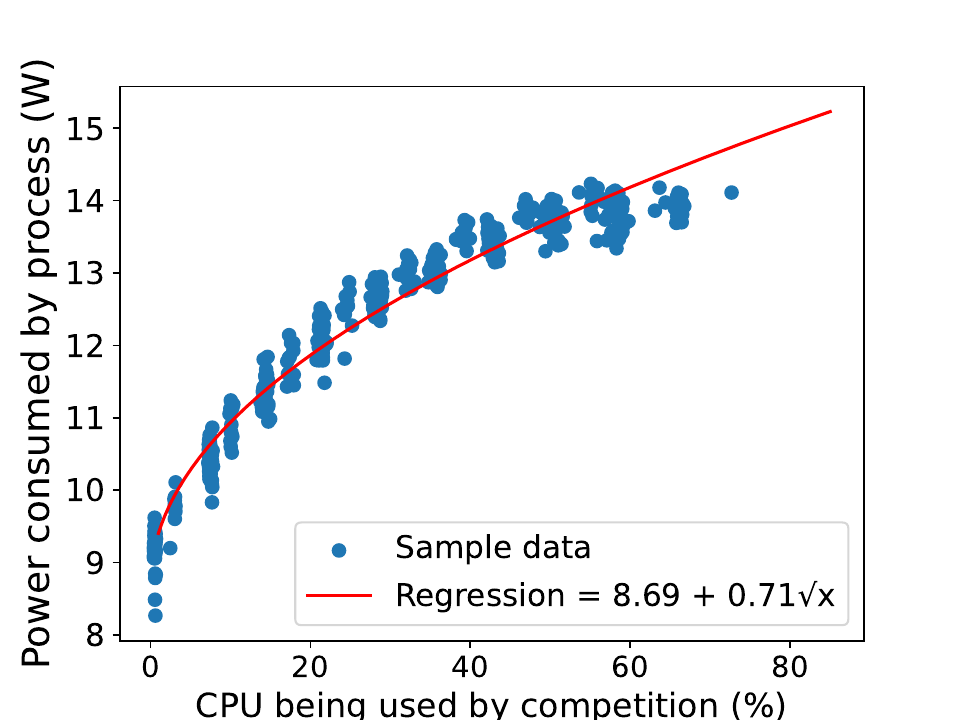}
    \caption{Worker 3 results}
    \Description{Graph showing the results for worker 3}
     \label{fig:w3}
  \end{minipage}
\hfill
\end{figure}
\begin{figure}
  \centering
  \begin{minipage}[b]{0.4\textwidth}
    \includegraphics[width=\textwidth]{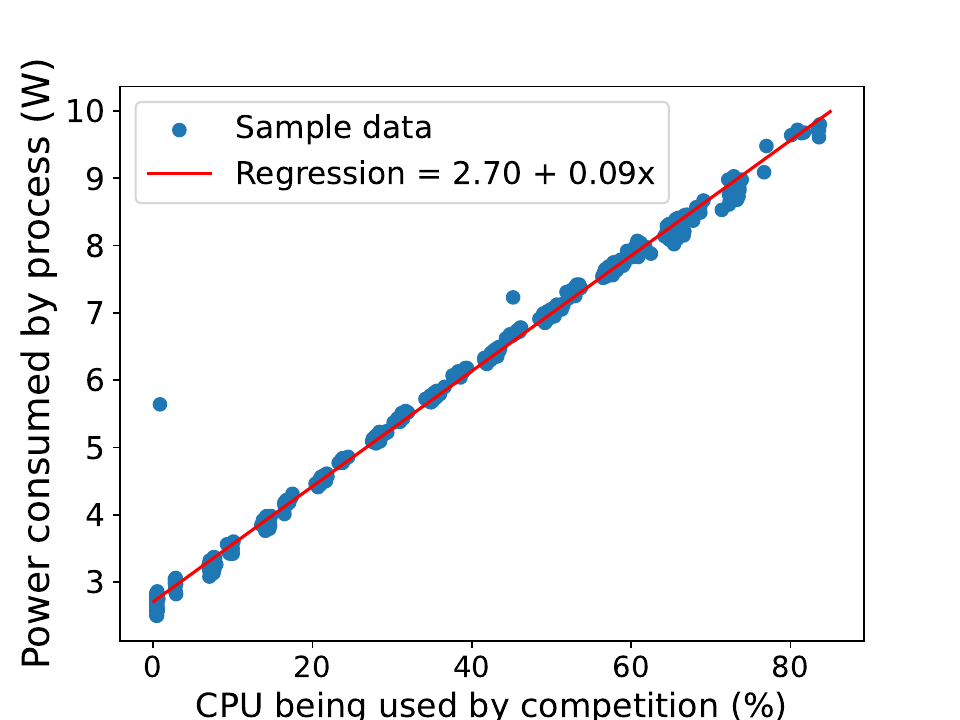}
    \caption{Worker 4 results}
    \Description{Graph showing the results for worker 4}
     \label{fig:w4}
  \end{minipage}
  \begin{minipage}[b]{0.4\textwidth}
    \includegraphics[width=\textwidth]{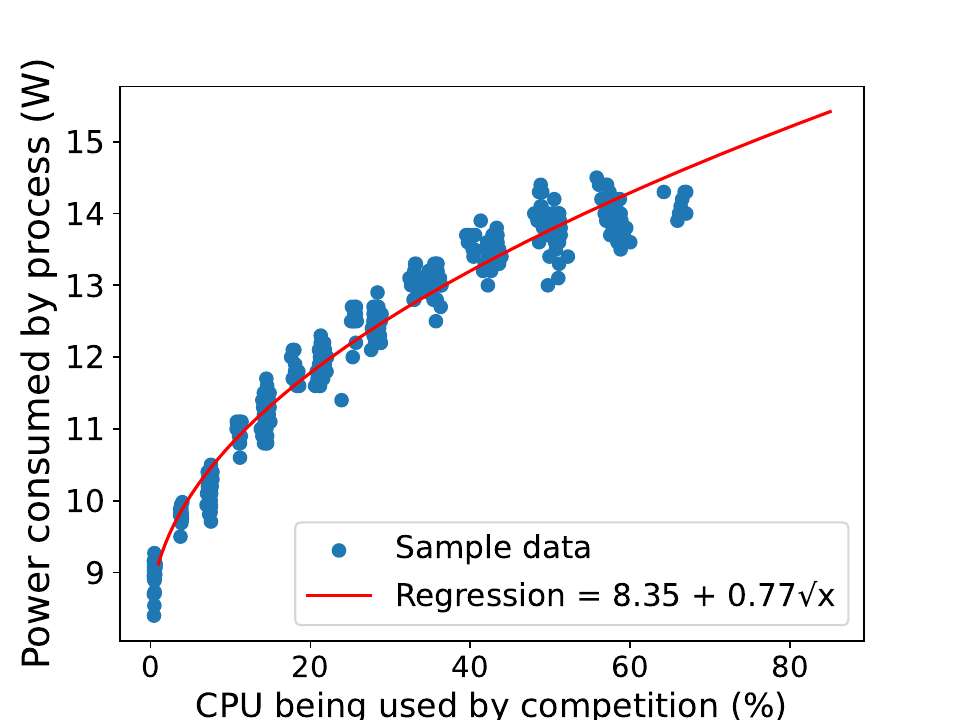}
    \caption{Worker 5 results}
    \Description{Graph showing the results for worker 5}
     \label{fig:w5}
  \end{minipage}
\end{figure}

After plotting the data, we proposed that the main process behavior could be represented by either a linear or an n-root function. 
For Workers 1 and 4 (Figure~\ref{fig:w1} and Figure~\ref{fig:w3}, respectively), the result was better described by a linear profile.
On the other hand, Workers 2 (Figure~\ref{fig:w2}), 3 (Figure~\ref{fig:w3}), and 5 (Figure~\ref{fig:w5}) were more accurately pictured by an n-root function.
Besides, it is known from Table~\ref{tab:hardware} that Workers 2, 3, and 5 have the same number of virtual cores on their CPUs (8), whilst Workers 1 and 4 have a lower number (4 and 6, respectively).
As there have been previous studies that proposed a correlation between the number of active threads and energy usage \cite{vcores}, we investigate the relationship between the number of virtual cores and the power consumption profile.

\subsection{Resource Competition Experiment - Gradual Increase with capped CPU}{\label{subsec:capped}}

Then, we suggested executing the same tests on 8-thread worker machines to investigate the effect of vCPUs. 
However, before running them, we would deactivate some of the cores to simulate a 4 or 6-thread computer and see if the function profile becomes linear.
The resulting dataset from these experiments is also publicly available in Zenodo \cite{campos}.

\begin{figure}[!bp]
  \centering
  \begin{minipage}[b]{0.4\textwidth}
    \includegraphics[width=\textwidth]{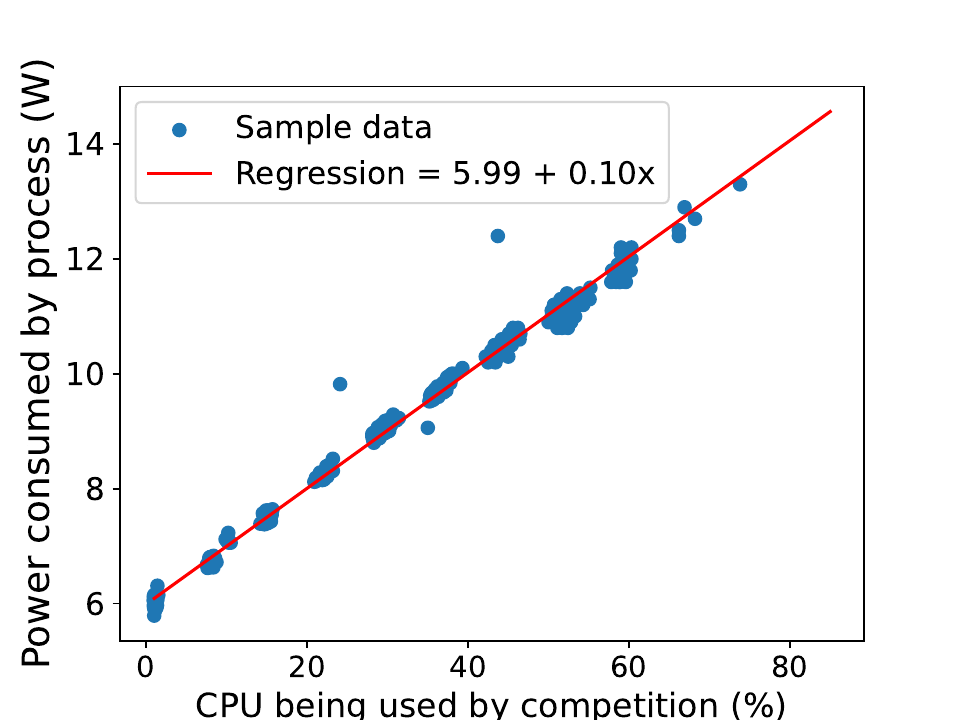}
    \caption{Worker 2 results with 4 cores}
    \Description{Graph showing the results for worker 2 with 4 cores}
    \label{fig:w24lin}
  \end{minipage}
\end{figure}

\begin{figure}
  \centering
  \begin{minipage}[b]{0.4\textwidth}
    \includegraphics[width=\textwidth]{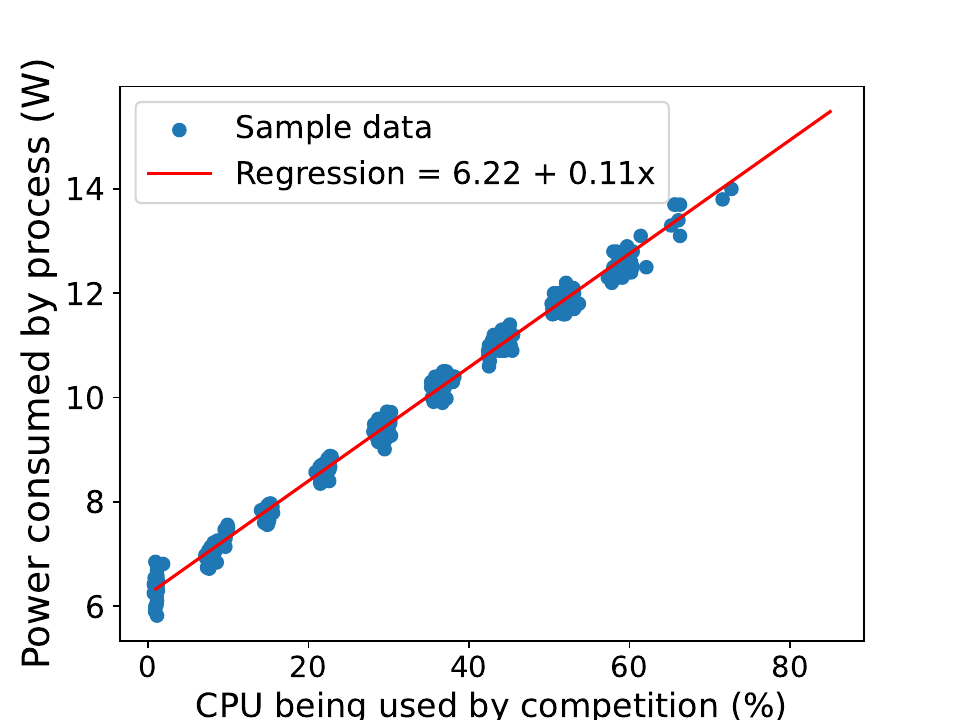}
    \caption{Worker 3 results with 4 cores}
    \Description{Graph showing the results for worker 3 with 4 cores}
    \label{fig:w34lin}
  \end{minipage}
 \hfill
  \begin{minipage}[b]{0.4\textwidth}
    \includegraphics[width=\textwidth]{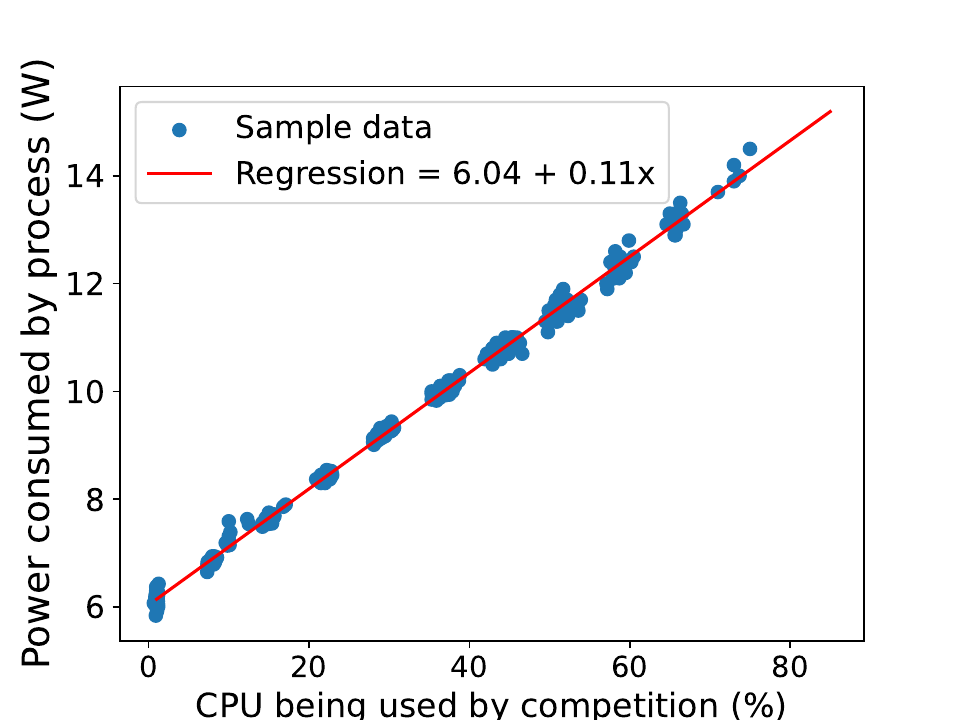}
    \caption{Worker 5 results with 4 cores}
    \Description{Graph showing the results for worker 5 with 4 cores}
    \label{fig:w54lin}
  \end{minipage}
  \hfill
  \begin{minipage}[b]{0.4\textwidth}
    \includegraphics[width=\textwidth]{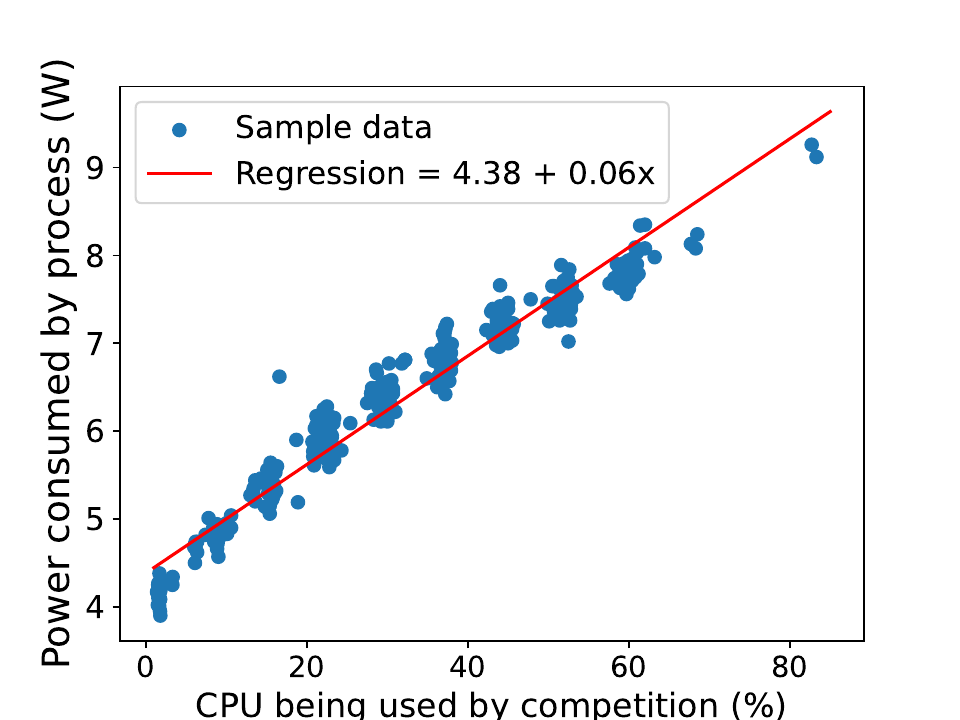}
    \caption{Worker 2 results with 6 cores and linear fit}
    \Description{Graph showing the results for worker 2 with 6 cores and linear fit}
    \label{fig:w26lin}
\end{minipage}
  \hfill
\end{figure}

When reducing the number of threads from 8 to 4 for the machines Worker 2 (Figure~\ref{fig:w24lin}), 3 (Figure~\ref{fig:w34lin}), and 5 (figure ~\ref{fig:w54lin}), we got a linear profile for $\mathcal{W}(p)$, just like the results for Worker 1 (the machine with 4 cores originally).
On the other hand, the results for capping the resources to 6 threads were mixed.

\begin{figure}
  \centering
  \begin{minipage}[b]{0.4\textwidth}
    \includegraphics[width=\textwidth]{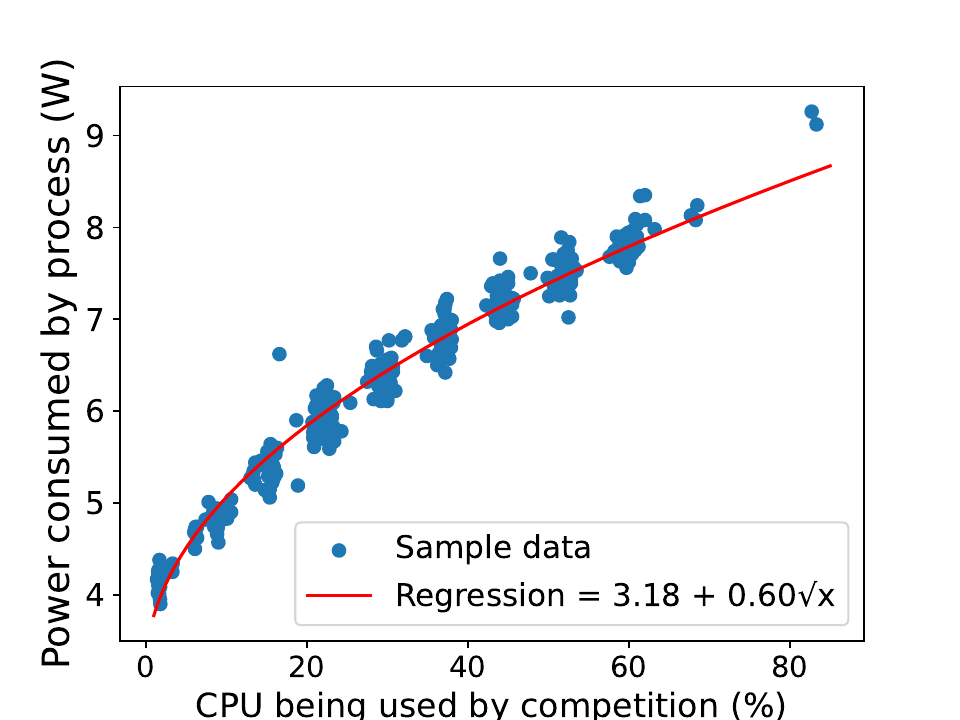}
    \caption{Worker 2 results with 6 cores and n-root fit}
    \Description{Graph showing the results for worker 2 with 6 cores and n-root fit}
    \label{fig:w26rt}
  \end{minipage}
  \begin{minipage}[b]{0.4\textwidth}
    \includegraphics[width=\textwidth]{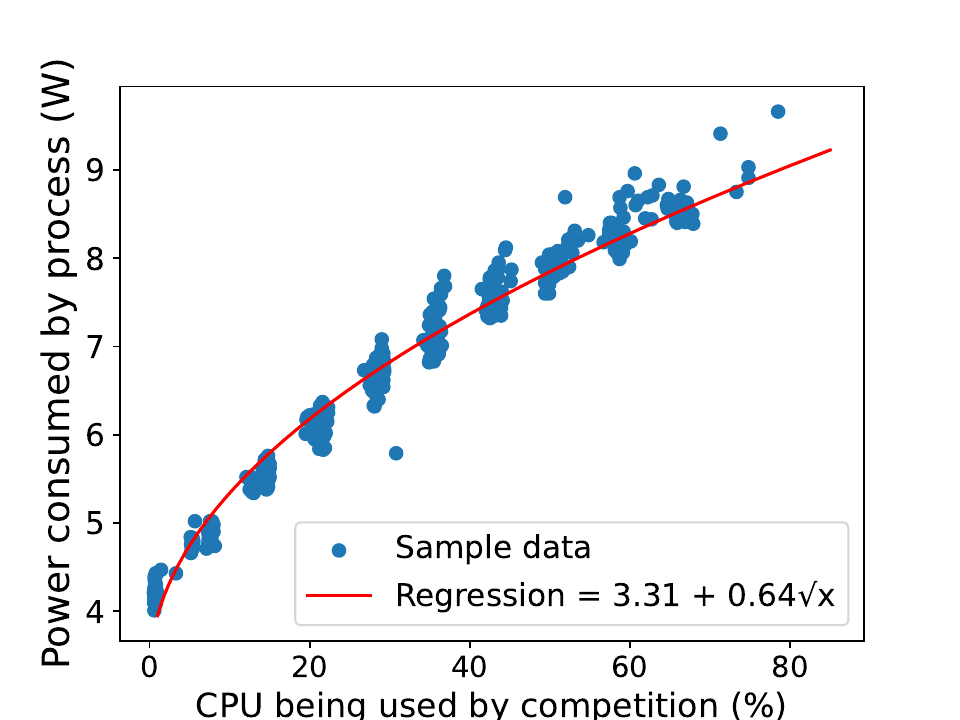}
    \caption{Worker 3 results with 6 cores and n-root fit}
    \Description{Graph showing the results for worker 3 with 6 cores and n-root fit}
    \label{fig:w36rt}
\end{minipage}
\hfill
  \begin{minipage}[b]{0.4\textwidth}
    \includegraphics[width=\textwidth]{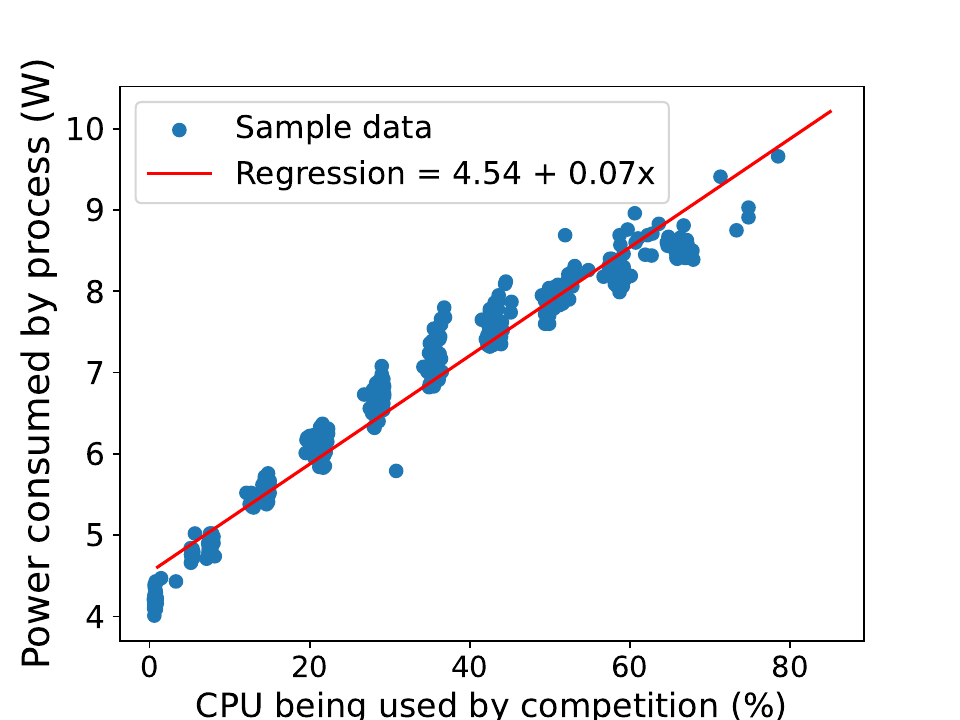}
    \caption{Worker 3 results with 6 cores and linear fit}
    \Description{Graph showing the results for worker 3 with 6 cores and linear fit}
    \label{fig:w36lin}
  \end{minipage}
\end{figure}

We executed this version of the test on Worker 2 and 3, and then, after gathering the data, we found out, as displayed in Figures~\ref{fig:w26rt} for Worker 2 and~\ref{fig:w36rt} for 3, that the best model for fitting this information was the n-root (differentiating from the original 6-core machine - Worker 4 - that provided a linear profile).
However, we also did linear fits, and the t-statistics were not significantly different from those with the n-root model (even though they were slightly worse).
This is graphically showcased by Figure~\ref{fig:w26lin} for Worker 2 and by Figure~\ref{fig:w36lin} for Worker 3.
In this scenario, we propose that there is a correlation between the number of virtual cores and the profile for $\mathcal{W}(p)$.

\subsection{Analysis \& Discussion}\label{subsec:analysis_and_discussion}
With the obtained data, it is possible to draw conclusions related to the $\mathcal{W}(p)$ profile and the machine's CPU.
First of all, for all machines with an n-root behavior, it is worth noting that not only was the function's type to represent the power consumption the same, but also the parameters were very similar (that is, $a$ and $b$ on $f(x) = a + b\sqrt[n]{x}$).
On Workers 2,3 and 5, the variation in $a$ did not surpass 7\%, while for $b$, the maximum difference reached the 10\% threshold.
With that, as these machines had significant differences in available RAM and disk space, it seems that the CPU type overrules any of these hardware variations since their function $\mathcal{W}(p)$ had approximately the same parameters.
This fact confirms the affirmation made in previous studies \cite{CPUpower} \cite{energySurvey} that the CPU can be considered the most relevant source of energy consumption on a machine.
Moreover, regarding the thread's discussion, we propose a correlation between the function profile and the number of virtual cores available.
In this case, we suggest that for lower numbers of cores, the $\mathcal{W}(p)$ behavior is linear (such as the results for Worker 1 and capped results for Workers 2,3 and 5 suggest).
As the number of these components increases, they gradually progress to an n-root behavior as they reach 6 cores (Worker 4 provided a linear pattern, whilst capped versions of Workers 2 and 3 had a mixed shape).
Then, as it reaches higher quantities of cores (up to 8, on our experimentation), the profile shifts completely to an n-root function (as Workers 2,3 and 5 results propose).
Following this line, it is possible to assume that most modern hardware would follow the latter version of $\mathcal{W}(p)$.

A conclusion taken from this is that the power consumption of a process behaves differently on each machine.
Then, when deciding where to instantiate a process (regarding attempting to use less energy), the fact that some machines have a linear or n-root profile brings intricacies when choosing the PC in which the process will dissipate less power.
To understand them, we shall propose an explanation of the possible $\mathcal{W}(p)$ derivatives.
When taking the linear version of the function ($\mathcal{W}_{lin}(p) = a + bp$), the derivative returns:
\begin{equation}\label{eq:lindWdp}
	\dot{\mathcal{W}_{lin}} = \frac{d\mathcal{W}_{lin}(p)}{dp}= b
\end{equation}
Given that $b$ is constant, the function increases at a constant rate.
That is, \textit{independent of how much CPU is being used by competing processes, the process energy consumption will increase with the same intensity}.
On the other hand, with the n-root model ($\mathcal{W}_{rt}(p) = c + d\sqrt[n]{p}$), the derivative is:

\begin{equation}\label{eq:rtdWdp}
	\dot{\mathcal{W}_{rt}} = \frac{d\mathcal{W}_{rt}(p)}{dp}= \frac{d}{n{p^{1-\frac{1}{n}}}}
\end{equation}

In this case, the derivative depends on $p$.
Moreover, it depends in a way that \textit{for lower values for $p$, the derivative is greater, whilst, for higher values, the derivative becomes tinier}.
To exemplify the relevance of these different behaviors, we propose a hypothetical scenario.
We assume that, for a certain process, $\mathcal{W}_{lin}(p_i) < \mathcal{W}_{rt}(p_i)$ for a certain $p_i \in [0,100 - q[$ and let $D(p) = \mathcal{W}_{lin}(p) - \mathcal{W}_{rt}(p)$.
Therefore, this shows $D(p_i) < 0$, and that means \textit{it is more energetically friendly to deploy the process on the machine with the linear profile}.
However, the derivative of $D(p)$ shows that (assume $k = 1 - \frac{1}{n}$):
\begin{equation}\label{eq:dDdp}
	\dot{D} = \frac{dD(p)}{dp}= b - \frac{d}{n{p^k}}
\end{equation}
\begin{equation}\label{eq:dDconc}
     \forall p > \sqrt[k]{\frac{d}{nb}},  \dot{D} > 0
\end{equation}
\begin{equation}\label{eq:dDconc2}
    \therefore \exists \, p_j > p_i \, |  \, D(p_j) = 0
\end{equation}
Thus, conclusion~\ref{eq:dDconc2} provides scenario ($p > p_j$) in which $D(p) > 0$. 
Therefore, if $p_j < 100 - q$, there will be an interval for $p$ (that is, $]p_j, 100 - q]$) for the competition \textit{in which it will be better (energy-wise) to use the machine with the n-root behavior rather than the linear one for the same $p$}.

This conclusion could be useful for virtualization scheduling scenarios. 
As previously detailed, when deciding the network topology for the instantiation of NS, the agent responsible for selecting the machine could use the $\mathcal{W}(p)$ models to assess which machine would receive the user's process in question.
Alongside performance metrics, the agent would be able to balance performance and energy consumption (with the model) to achieve the necessary Quality of Service without hurting the environment as much.

Besides, such results could be used in more generalized virtualization scenarios, such as pricing the utilization of cloud resources.
Instead of considering only the type of hardware that is being used, cloud companies could also assess how much energy the user's processes are consuming.
In this case, enterprises could more accurately evaluate how many resources are being consumed by the user, which would lead to a value that would represent the usage better than only considering hardware.


\section{Considerations \& Future work}{\label{sec:cons}}

In the resource competition context, we evaluated the state of the art.
We learned that the relationship between energy consumption and CPU usage at the process level was not yet consolidated.
Thus, we applied an empirical method that, given its results and analysis, successfully establishes a function $\mathcal{W}(p)$ to represent the energy consumed by a process.

The conducted experiments showed that the profile of a process's energy consumption as a function of the CPU used by the competing processes is dependent on the number of CPU virtual cores.
When there are 4 threads, that relationship is linear.
For 6 cores, a transition between linear and n-root is seen.
Then, when 8 threads are available, the behavior seen is of an n-root.
Our experiments suggest that, for a higher number of threads, the n-root behavior remains.

In this study, we observe that a lower number of cores means a linear relationship between energy consumption and competition CPU usage, meaning the rate of energy consumption growth matches the rate of CPU usage growth.
However, a higher number of cores leads to an n-root behavior for that relationship.
This means that at low levels of CPU usage, energy consumption grows at a higher rate, whereas at high levels of CPU usage, energy consumption grows at a lower rate.
This leads to the conclusion that, when the competition is at low CPU usage, it is less energy-consuming to run a process on a machine with fewer cores.
On the other hand, when the competition is at high CPU usage, it consumes less energy to run a process on a machine with more cores.
These results significantly impact resource allocation and pricing in cloud services, as competition should be a factor considered when allocating new resources to minimize energy consumption and when charging users.

In future work, we aim to extend our analysis to machines with a higher number of CPU threads to further validate the n-root behavior observed.
Additionally, extending our research to include \acp{VM} would be highly beneficial.
By measuring and modeling the energy consumption profile of a process as a function of the competition it faces within VMs, we could uncover new insights that enhance our understanding of resource management or pricing in virtualized environments.


\begin{acks}
The authors acknowledge FAPESP for supporting the thematic project SFI2 - Slicing Future Internet Infrastructures (2018/23097-3 - MCTIC/CGI), as well as the Scientific Initiation proposals 2023/13381-4 and 2023/13383-7. We also thank the ANIMA Institute and CNPq (140303/2021-9) for research scholarships. Finally, we are grateful to UDESC and the LabP2D laboratory for their partnership.
\end{acks}



\begin{acronym}[XGBoost]
\acro{3GPP}{3rd Generation Partnership Project}
\acro{6G}{sixth-generation}
\acro{5G}{fifth-generation}
\acro{A2C}{Advantage Actor-Critic}
\acro{ACM}{Association for Computing Machinery}
\acro{AN}[AN]{Access Network}
\acro{AGNS}{Automatic Generation of Network Slices}
\acro{AI}{Artificial Intelligence}
\acro{AP}{Access Point}
\acro{ARIMA}{Autoregressive Integrated Moving Average}
\acro{B2C}[B2C]{Business to consumer}
\acro{B2B}[B2B]{Business to business}
\acro{B5G}{Beyond 5G}
\acro{BE}{Best Effort}
\acro{BS}{Base Station}
\acro{C-RAN}{Cloud-RAN}
\acro{CBDA}{Chain-Based Data Aggregation}
\acro{CC}{Cognitive Cycles}
\acro{CN}{Core Network}
\acro{CNN}{Convolutional Neural Network}
\acro{CPU}{Central Processing Unit}
\acro{CSI}{Channel State Information}
\acro{CT}{Communication Technology}
\acro{CU}{Central Unit}
\acro{DCIE}{Data Center Infrastructure Efficiency}
\acro{DCMAB}{Deep Contextual MAB}
\acro{DDPG}{Deep Deterministic Policy Gradient}
\acro{DDQN}{Double Deep Q-Network}
\acro{DQL}{Deep Q-Learning}
\acro{DQN}{Deep Q-Network}
\acro{DL}{Deep Learning}
\acro{DLNN}{Deep Learning Neural Network}
\acro{DRL}{Deep Reinforcement Learning}
\acro{DNN}{Deep Neural Network} 
\acro{DU}{Distributed Unit}
\acro{DPI}{Deep Packet Inspection}
\acro{E2E}{end-to-end}
\acro{EC}{Edge Controller}
\acro{EE}{Energy Efficiency}
\acro{eMBB}{Enhanced Mobile Broadband}
\acro{EPDA}{Exponential Power Descent Algorithm}
\acro{ETS}{Exponential Smoothing}
\acro{ETSI}{European Telecommunications Standards Institute}
\acro{FL}{Federated Learning}
\acro{F-AP}{Fog Radio Access Point}
\acro{F-RAN}{Fog-RAN}
\acro{F-UE}{Fog-UE}
\acro{GBR}{Gradient Boosting Regressor}
\acro{GCN}{Graph Convolutional Network}
\acro{GHG}{greenhouse gas}
\acro{H2H}{Human-to-Human}
\acro{HDBSCAN}{Hierarchical Density Based Spatial Clustering}
\acro{HetNet}{Heterogeneous Network}
\acro{HW}{Holt-Winters}
\acro{ICT}{Information and Communication Technology}
\acro{IEEE}{Institute of Electrical and Electronics Engineers}
\acro{IETF}{Internet Engineering Task Force}
\acro{ILP}{Integer Linear Programming}
\acro{IP}{Internet Protocol}
\acro{InP}{Infrastructure Provider}
\acro{IoT}{Internet of Things}
\acro{IIoT}{Industrial Internet of Things}
\acro{ISP}{Internet Service Provider}
\acro{ITU-T}{International Telecommunication Union Telecommunication Standardization Sector}
\acro{KPI}{Key Performance Indicator}
\acro{LCM}{Life Cycle Management}
\acro{LIME}{Local Interpretable Model-Agnostic Explanations}
\acro{LR}{Logistic Regression}
\acro{LSTM}{Long Short Term Memory}
\acro{LTE}{Long Term Evolution}
\acro{M2M}{Machine-to-Machine}
\acro{MAB}{Multi-Armed Bandit}
\acro{MANO}{Management and Orchestration}
\acro{MEC}{Multi-access Edge Computing}
\acro{MDP}{Markov Decision Process}
\acro{mIoT}{Massive IoT}
\acro{mMTC}{Machine Type Communication}
\acro{MPLS}{Multiprotocol Label Switching}
\acro{ML}{Machine Learning}
\acro{MVNO}{Mobile Virtual Network Operator}
\acro{N3AC}{Neural Network Admission Control}
\acro{NECOS}{Novel Enablers for Cloud Slicing}
\acro{NFV}{Network Function Virtualization}
\acro{NGMN}{Next-Generation Mobile Networks}
\acro{NLP}{Natural Language Processing}
\acro{NN}{Neural Network}
\acro{NS}{Network Slicing}
\acro{NSaaS}{Network Slice-as-a-Service}
\acro{NSMF}{Network Slice Management Function}
\acro{NSS}{Network Slice Subnet}
\acro{NS-3}{Network Simulator 3}
\acro{NSP}{Network Service Provider} 
\acro{ONETS}{Online NETwork Slice Broker}
\acro{O-RAN}{Open RAN}
\acro{PI}{Permutation Importance}
\acro{PRB}{Physical Resource Block}
\acro{PoP}{Point of Presence}
\acro{PT}{Prospect Theory}
\acro{PUE}{Power Usage Effectiveness}
\acro{QCI}{QoS Class Identifier}
\acro{QL}{Q-Learning}
\acro{QoE}{Quality of Experience}
\acro{QoT}{Quality of Transmission}
\acro{QoS}{Quality of Service}
\acro{RIC}{RAN Intelligent Controller}
\acro{SFC}{service function chain}
\acro{UAV}{Unmanned Aerial Vehicle}
\acro{UE}{User Equipment}
\acro{uRLLC}{ultra-Reliable and Low-Latency Communications}
\acro{RAM}{Random Access Memory}
\acro{RAN}{Radio Access Network}
\acro{RAT}{Radio Access Technology}
\acro{RF}{Random Forest}
\acro{RFR}{Random Forest Regressor}
\acro{RL}{Reinforcement Learning}
\acro{RU}{Radio Unit}
\acro{SARSA}{State-Action-Reward-State-Action}
\acro{SDN}{Software-Defined Network}
\acro{SFI2}{Slicing Future Internet Infrastructures}
\acro{SLA}{Service Level Agreement}
\acro{SLAW}{Self-similar least-action human walk}
\acro{SDO}{Standards Developing Organization}
\acro{SMDP}{Semi-Markov Decision Process}
\acro{SON}{Self-Organized Network}
\acro{SHAP}{SHapely Additive Explanations}
\acro{SL}{Supervised Learning}
\acro{SMO}{Service Management and Orchestration}
\acro{SP}{Slice Provider}
\acro{SVM}{Support Vector Machine}
\acro{SBLR}{Sparse Bayesian Linear Regression}
\acro{TL}{Transfer Learning}
\acro{TN}{Transport Network}
\acro{UCB}{Upper Confidence Bound}  
\acro{UE}[UE]{User Equipment}
\acro{UL}{Unsupervised Learning}
\acro{V2I}{Vehicle-to-Infrastructure}
\acro{V2N}{Vehicle-to-Network}
\acro{V2P}{Vehicle-to-Pedestrian}
\acro{V2V}{Vehicle-to-Vehicle}
\acro{V2X}{Vehicle-to-Everything}
\acro{vBS}{virtual Base Station}
\acro{VM}{Virtual Machine}
\acro{VNE}{Virtual Network Embedding}
\acro{VNF}[VNF]{Virtualized Network Function}
\acro{VNFD}[VNFD]{Virtualized Network Function Descriptor}
\acro{VS}{Virtual Slicing}
\acro{WET}{Wireless Power Transmission}
\acro{Wi-Fi}{Wireless Fidelity}
\acro{XAI}[XAI]{eXplainable Artificial Intelligence}
\acro{XGBoost}{Extreme Gradient Boosting}
\acro{ZSM}{Zero touch network \& Service Management}

\end{acronym}

\bibliographystyle{ACM-Reference-Format}
\bibliography{References/references.bib}

\end{document}